*healthcare*

# A Deep Learning Approach for Brain Tumor Classification and Segmentation Using a Multiscale Convolutional Neural Network

Francisco Javier Díaz-Pernas, Mario Martínez-Zarzuela 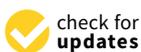, Míriam Antón-Rodríguez and David González-Ortega *

Department of Signal Theory, Communications and Telematics Engineering, Telecommunications Engineering School, University of Valladolid, 47011 Valladolid, Spain; pacper@tel.uva.es (F.J.D.-P.); marmar@tel.uva.es (M.M.-Z.); mirant@tel.uva.es (M.A.-R.)
* Correspondence: davgon@tel.uva.es; Tel.: +34-983-423-000 (ext. 5552)

**Abstract:** In this paper, we present a fully automatic brain tumor segmentation and classification model using a Deep Convolutional Neural Network that includes a multiscale approach. One of the differences of our proposal with respect to previous works is that input images are processed in three spatial scales along different processing pathways. This mechanism is inspired in the inherent operation of the Human Visual System. The proposed neural model can analyze MRI images containing three types of tumors: meningioma, glioma, and pituitary tumor, over sagittal, coronal, and axial views and does not need preprocessing of input images to remove skull or vertebral column parts in advance. The performance of our method on a publicly available MRI image dataset of 3064 slices from 233 patients is compared with previously classical machine learning and deep learning published methods. In the comparison, our method remarkably obtained a tumor classification accuracy of 0.973, higher than the other approaches using the same database.

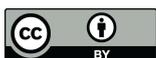

## 1. Introduction

Automatic segmentation and classification of medical images play an important role in diagnostics, growth prediction, and treatment of brain tumors. An early tumor brain diagnosis implies a faster response in treatment, which helps to improve patients' survival rate. Location and classification of brain tumors in large medical images databases, taken in routine clinical tasks by manual procedures, have a high cost both in effort and time. An automatic detection, location, and classification procedure is desirable and worthwhile [1].

There are several medical imaging techniques used to acquire information about tumors (tumor type, shape, size, location, etc.), which are needed for their diagnosis [2]. The most important techniques are Computed Tomography (CT), Single-Photon-Emission Computed Tomography (SPECT), Positron Emission Tomography (PET), Magnetic Resonance Spectroscopy (MRS), and Magnetic Resonance Imaging (MRI). These techniques can be combined to obtain more detailed information about tumors. Anyhow, MRI is the most used technique due to its advantageous characteristics. In MRI acquisition, the scan provides hundreds of 2D image slices with high soft tissue contrast using no ionizing radiation [2]. There are four MRI modalities used in diagnosis: T1-weighted MRI (T1), T2-weighted MRI (T2), T1-weighted contrast-enhanced MRI (T1-CE), and Fluid Attenuated Inversion Recovery (FLAIR). Each MRI modality produces images with different tissue contrasts; thus, some are more suited to search a specific kind of tissue than others. T1 modality is typically used to work with healthy tissues. T2 images are more appropriate to detect borders of edema regions. T1-CE images highlight tumor borders and FLAIR images favor the detection of edema regions in Cerebrospinal Fluid [3]. Provided that the goal of MRI image processing is to locate and classify brain tumors, T1-CE modality is adequate and, as it is shown in this paper, sufficient.

In the last decades, Brain Tumor Imaging (BTI) has grown at an exponential rate. More specifically, the number of works about brain tumor quantification based on MRI images has increased significantly [2]. Brain Tumor Segmentation (BTS) consists in differentiating tumor-infected tissues from healthy ones. In many BTS applications, the brain tumor image segmentation is achieved by classifying pixels, thus the segmentation problem turns into a classification [4].

The aim of the work presented in this paper is to develop and test a Deep Learning approach for brain tumor classification and segmentation using a Multiscale Convolutional Neural Network. To train and test the proposed neural model, a T1-CE MRI image dataset from 233 patients, including meningiomas, gliomas, and pituitary tumors in the common views (sagittal, coronal, and axial), has been used [5]. Figure 1 shows examples of these three types of tumors. Additional information on the dataset is included in Section 2.2. Our model is able to segment and predict the pathological type of the three kinds of brain tumors, outperforming previous studies using the same dataset.

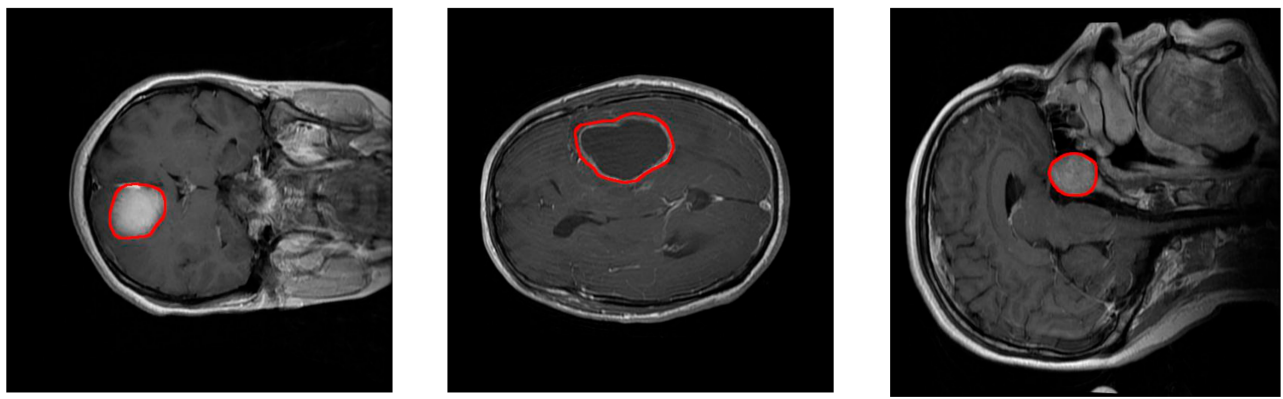

**Figure 1.** Examples of MRI images of the T1-CE MRI image dataset. Left: coronal view of a meningioma tumor. Center: Axial view of a glioma tumor. Right: sagittal view of a pituitary tumor. Tumor borders have been highlighted in red.

In the BTS field, two main tumor segmentation approaches can be found: generative and discriminative. Generative approaches use explicit anatomical models to obtain the segmentation, while discriminative methods learn image features and their relations using gold standard expert segmentations [2]. Published studies following the discriminative approach have evolved from using classical Machine Learning [6–9] to more recent Deep Learning techniques [10–14].

In works using classical Machine Learning techniques, the segmentation pipeline includes a preprocessing stage designed for feature extraction. For example, in Sachdeva et al. [6], a semi-automatic system obtained the tumor contour, and 71 features were then computed using the intensity profile, the co-occurrence matrix, and Gabor functions. In addition, skull stripping is a common preprocessing stage in classical discriminative approaches, although it presents drawbacks such as the selection of parameters or the need of prior information about the images and a high computational time [15]. The extracted features are used as the input for a classification or a segmentation stage. In the aforementioned work [6], two classifiers, SVM (Support Vector Machine) and ANN (Artificial Neural Network), were compared. The obtained accuracy ranged from 79.3% to 91.7% for SVM and from 75.6% to 94.9% for ANN. Analogously, Sharma and Chhabra [7] proposed an automatic brain tumor classification system using 10 features and a Back Propagation Network as the classifier, and they reached an accuracy of 95.3%. Iftekharuddin et al. [8] explored using fractal wavelet features as input to a SOM (Self Organizing Map) classifier and achieved an average precision of 90%. Other examples of tumor classification pipelines use instance-based learning: Havaei et al. [9] developed a semi-automatic system based on a kNN classifier. They used the well-known BRATS 2013 dataset and obtained the second-best score over the complete and core tests, reporting Dice similarities of 0.85 and 0.75 for complete and core tumor regions, respectively.

In the last years, the next step in the evolution of machine learning techniques has been to allow computers to discover the features that optimally represent the data. This concept is the foundation of Deep Neural Networks [4], which transform the problems to solve from feature-driven into data-driven. Within deep neural networks, Convolutional Neural Networks (CNN) [16] and Fully Convolutional Networks (FCN) [17,18] have been applied to a broad range of applications. In particular, they are widely used today for generic image processing [19] and specifically in medical image analysis [1,4,20].

The advent of Deep Learning has had a tremendous impact in the design of medical imaging systems, and newer techniques have been widely adopted by the researchers over the last few years [21]. The advances in the field have opened a very exciting time for collaboration of machine learning scientists and radiologists [22]. Havaei et al. [10] explored the use of a fully convolutional system for the BRATS 2013 dataset, just after their kNN classifier proposal [9], commented before. Most of the recent brain tumor classification studies using Deep Learning are based on CNNs or FCNs. Moeskops et al. [11] used a multiscale CNN to segment brain tissues and white matter hyperintensities. They used images from the MRBrainS13 challenge [12] to evaluate their CNN architecture and reached a precision of about 85%. One of the most significant advances in the field was achieved by Ronneberger et al. [13], who proposed an interesting CNN named U-net. They used this CNN architecture to segment neural structures in electron microscopy images from the ISBI challenge [14]. U-net is composed by 5 convolutional stages and 5 deconvolutional stages shaping a U, from where it takes its name. They obtained the best results in the ISBI challenge [14].

However, all the previously cited studies have in common that they aimed to segment areas belonging to brain tumors without the need to also classify these areas as belonging to different brain tumor types. As explained above, in our study the objective was to develop a new Deep Learning model for both tumor segmentation and classification. A relevant previous related work is the study of Mohsen et al. [23], who proposed a deep learning classifier combined with a discrete wavelet transform (DWT) and principal components analysis (PCA) to classify a dataset with three different brain tumors. Four other deep learning relevant studies following the same purpose have in common the use of the same dataset that we adopted in this work, which is important to contrast and endorse the performance results of the proposed model presented here. Abiwinanda et al. [24] proposed a two-layer CNN architecture. Pashaei et al. [25] proposed two different methods: in the first method, a CNN model was used for classification, and in the second method, the features extracted by CNN were the inputs of a KELM method. The KELM method is a learning algorithm that consists of layers of hidden nodes. Sultan et al. [26] proposed a CNN network including 16 convolution layers. Finally, Anaraki et al. [27] presented a hybrid method combining the use of CNNs and genetic algorithm (GA) criteria to improve the network architecture.

In this paper, we present a fully automatic CNN-based segmentation and classification method of MRI images for three types of tumor: meningioma, glioma, and pituitary tumors. Unlike previous CNN-based works [10,23–27], our neural network includes three processing pathways to deal with the information in three spatial scales. Previous research conducted by our research group and inspired in the inherent multiscale operation of the Human Visual System (HVS) [28] led us to propose this multiscale processing strategy for tumor classification on the assumption that a multiscale approach would be effective extracting discriminant texture features of different kinds of tumors. The proposed model is very coherent with the HVS processing. In visual stimuli processing, the HVS works in two main modes: pre-attentive and attentive vision. Pre-attentive mode is instantaneous and parallel, covering wide areas of the vision field, while attentive processing acts over limited regions of the visual field, establishing a serial search through focused attention [29]. In this processing, simple cells in the V1 area of the visual cortex in the brain filter the stimuli coming from the LGN (Lateral Geniculate Nucleus), through frequency and bandwidth filtering. After that filtering, the processed features are concatenated in the inferotemporal area during the cognitive process [30]. In addition, our method uses a sliding window technique differing from U-net [13] and other methods, in which the full image is used as the input.

The rest of the paper is organized as follows: Section 2 introduces the proposed CNN model, describes the adopted dataset and how it was used for training and measuring the performance of the neural network. Later, Section 3 presents the experimental results and discusses these results including a performance comparison with other methods. Finally, Section 4 draws the main conclusions about the presented work.

## 2. Materials and Methods

### 2.1. Proposed Convolutional Neural Network and Implementation Details

In this paper, we propose a multi-pathway CNN architecture (see Figure 2) for tumor segmentation. The CNN architecture processes an MRI image (slice) pixel by pixel covering the entire image and classifying each pixel using one of four possible output labels: 0—healthy region, 1—meningioma tumor, 2—glioma tumor, and 3—pituitary tumor.

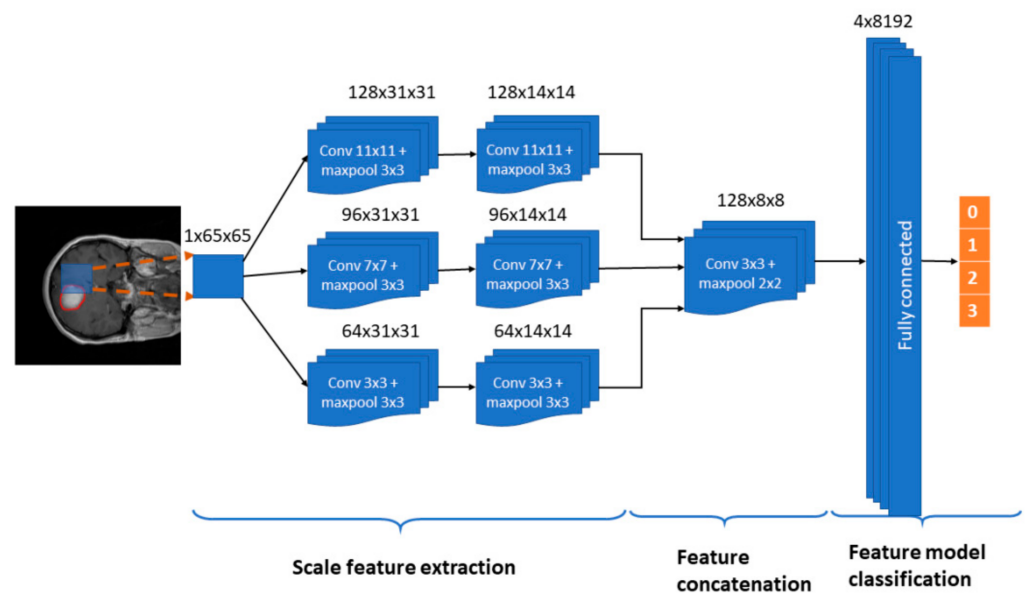

**Figure 2.** The proposed Convolutional Neural Networks (CNN) architecture. Input: $1 \times 65 \times 65$ sliding windows. Model: Three pathways (large, medium, and small feature scales) with 2 convolutional layers and max-pooling, a convolutional layer with concatenation of the three pathways, and a fully connected stage that leads to a classification in one out of the four possible output labels: 0—healthy region, 1—meningioma tumor, 2—glioma tumor, and 3—pituitary tumor. A dropout mechanism between the concatenation and fully connected stages is included.

In our approach, we use a sliding window, thus each pixel is classified according to a $N \times N$ neighborhood or window, which is the input to our CNN architecture (see Figure 2). Every window is processed through three convolutional pathways with three scale (large, medium, and small) kernels that extract the features. In our implementation, we chose a window of $65 \times 65$ pixels and kernels of size $11 \times 11$, $7 \times 7$ and $3 \times 3$ pixels, respectively. The decision on the size of the windows was taken after preliminary configuration tests in which also $33 \times 33$ px and $75 \times 75$ px window sizes were tested.

Each pathway is composed by two convolutional stages with ReLU rectification and a $3 \times 3$ max-pooling kernel with a stride value of 2. The number of feature maps in the large, medium, and small pathways is 128, 96, and 64, respectively. We propose the use of a larger number of maps for larger scales with the assumption that the window features extracted using different filtering scales help to define the three kinds of tumor to be classified.

Scale features from the three pathways are concatenated in a convolutional layer with a $3 \times 3$ kernel with a ReLU activation function and a $2 \times 2$ max-pooling kernel with a stride value of 2. The output of this stage enters a fully connected stage where 8192 concatenated scale features compose the classification method towards the four prediction label types. In order to prevent overfitting, the model includes a dropout layer before the fully connected layer. The last layer uses a softmax activation function.

The proposed CNN has been implemented using Pytorch™. The number of trainable parameters in the neural network is near three million (2,856,932). All the tests have been performed in a Linux environment with an Intel Core i7 CPU and an Nvidia GTX1080 T1-11GB GPU. The training process took 5 days, and the average prediction time per image was 57.5 s.

### 2.2. Dataset and Data Preprocessing

In clinical settings, usually only a certain number of slices of brain CE-MRI with a large slice gap, not a 3D volume, are acquired and available. It is difficult to build a 3D model with such sparse data. Hence, the proposed method is based on 2D slices collected from 233 patients, acquired from Nanfang Hospital, Guangzhou, China, and General Hospital, Tianjing Medical University, China, from 2005 to 2010 [5].

This dataset contains 3064 slices and includes meningiomas (708 slices), gliomas (1426 slices), and pituitary tumors (930 slices) in the common views (sagittal, coronal, and axial). Figure 1 shows examples of these three types of tumors. This dataset provides also 5-fold cross-validation indices. By using this information, 80% (2452) of the images are employed for training and 20% (612 images) are used for obtaining performance measurements. The process is repeated 5 times.

The images have an in-plane resolution of $512 \times 512$ pixels with pixel size $0.49 \times 0.49$ mm$^2$. The slice thickness is 6 mm, and the slice gap is 1 mm. The tumor border was manually delineated by three experienced radiologists. Every slice in the dataset has an information structure attached containing the patient's pid; the tumor type label ($l_{gt}$): 1 for meningioma, 2 for glioma, and 3 for pituitary tumor; the coordinates vector (***x,y***) of the points belonging to the tumor border, and the tumor mask (*Tij*): a binary image where the tumor positions contain a value of 1 and the healthy ones a value of 0. The pair $\{l_{gt}, Tij\}$ will be the ground truth in the training process.

During training, data augmentation using an elastic transform [31] has been used to prevent overfitting of the neural network. Figure 3 shows an example of this transformation. Data augmentation procedure doubled the number of training images available on each fold iteration up to 4904 images. A thorough process was conducted to extract $65 \times 65$ px training examples from every image in the training dataset: 150 true positive window examples and 325 true negative window examples per tumor. The pixels on these windows were scaled using pixel standardization (zero mean and unit variance) across the entire training dataset.

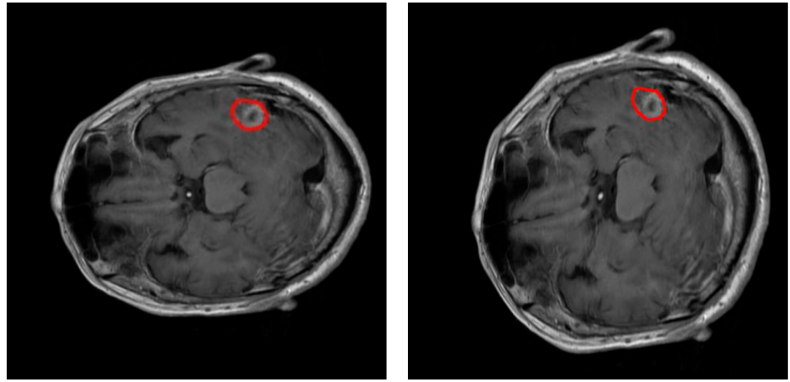

**Figure 3.** Example of elastic transformation used in the data augmentation. Left: original slice. Right: image transformed. In both images, the edges of the tumor have been highlighted in red.

### 2.3. Neural Network Training and Performance Measurements

The proposed CNN and performance measurements were obtained after a 5-fold cross validation technique with the slice train/test subgroups specified in the dataset, as explained before.

On each fold iteration, the proposed model was trained for a total of 80 epochs, using a Stochastic Gradient Descent (SGD) optimizer with a starting learning rate of 0.005 and a momentum coefficient of 0.9. In addition, we used a learning rate exponential decay every 20 epochs. The dropout parameter was set to 0.5.

Within each fold, performance of the model was tested over 612 images, using a sliding window approach: every testing window of size $65 \times 65$ px enters the trained CNN so that a predicted tumor type label $l_p$, is obtained. Segmentation of the tumor is achieved considering the region containing all the pixels identified as a given tumor type. The previously stored global mean and standard deviation of pixels in the training windows, computed for data preprocessing, are used to normalize the testing windows before they enter the CNN.

Once all pixels $(i, j)$ from the input slice are labeled, $P_{ij}$ (see Equation (1)), the classification function can be calculated using Equation (2). The predicted label $l_p$, which indicates the tumor type in a slice, is determined from vector $\{f_l,\ l = 1, 2, 3\}$ in the classification function. The classification function calculates the relation between sizes of the predicted label, $l$, $\{P_{ij} == l\}$ (number of pixels with the predicted label $l$), and the complete prediction, $\{P_{ij} > 0\}$ (number of pixels with a predicted label belonging to any tumor). The predicted label, $l_p$, will belong to the label with the greatest relation in size that is above the minimal relation defined by the confidence threshold, $\tau_c \in [0, 1]$.

$$P_{ij}, \begin{cases} P_{ij}, = 0, & if\ (i, j)\ is\ healthy\ position \\ P_{ij}, = 1, & if\ (i, j)\ is\ meningioma\ tumor \\ P_{ij}, = 2, & if\ (i, j)\ is\ glioma\ tumor \\ P_{ij}, = 3, & if\ (i, j)\ is\ pituitary\ tumor \end{cases} \tag{1}$$

$$f_l = \begin{cases} \frac{|P_{ij} == l|}{|P_{ij} > 0|} > \tau_c \\ 0 \end{cases} \tag{2}$$

$$l_p = \begin{cases} argmax\{f_l > 0\} & if\ \{f_l > 0\} \neq \{\varnothing\} \\ -1 & nonclassified \end{cases} \tag{3}$$

where $|.|$ is the size or pixel number (number of pixels in the slice that meet the conditions inside).

For performance measurements, we compute the Confusion Matrix, together with the Dice and Sensitivity scores [3]. In addition, with the aim of evaluating the precision of the predicted tumor type, we defined the predicted tumor type ratio score, *pttas*. This index calculates the relation in size between the correctly labeled tumor regions $\{P_{ij} = l_{gt}\}$ and all the tumor regions determined by our methodology $\{P_{ij} > 0\}$. In the *pttas* index, the ratio is measured over the total size of the predicted tumor, while in the Sensitivity index the ratio is calculated over the size of the ground truth tumor.

The evaluation metrics are calculated using the following Equations:

$$Dice(P, T) = \frac{|P_1 \wedge T_1|}{(|P_1| + |T_1|)/2} = \frac{2\,TP}{2\,TP + FP + FN} \tag{4}$$

$$Sensitivity(P, T) = \frac{|P_1 \wedge T_1|}{|T_1|} = \frac{TP}{TP + FN} \tag{5}$$

$$pttas = \frac{|P_1|}{|P > 0|} \tag{6}$$

where $TP$ is the number of true positives, $FP$ is the number of false positives, $FN$ is the number of false negatives, $\wedge$ is the logical AND operation, $|.|$ is the size or pixel number (number of pixels in the slice that meet the conditions inside), $P_1$ represents the addition of $TP$ and $FP$ (the predicted positives $\{P_{ij} = l_{gt}\}$), and $T_1$ represents the addition of $TP$ and $FN$ $\{T_{ij} = 1\}$.

## 3. Results and Discussion

As explained in the previous section, the neural network was tested using the 5-fold cross validation train/test subgroups specified in the dataset. The quantitative results obtained are displayed in the figures and tables below in this section.

Figure 4 shows the performance of our method for three slices. Figure 4—left—corresponds to a slice with a meningioma tumor, Figure 4—center—shows a glioma tumor, and Figure 4—right—a pituitary tumor. The images show the tumor segmentation obtained, $\{P_{ij} > 0\}$. To mark them with different colors, color images were generated where the predicted tumor region is filled with red color, the ground truth tumor region, $T_{ij}$, is colored green, and the intersection of those two regions appears in yellow.

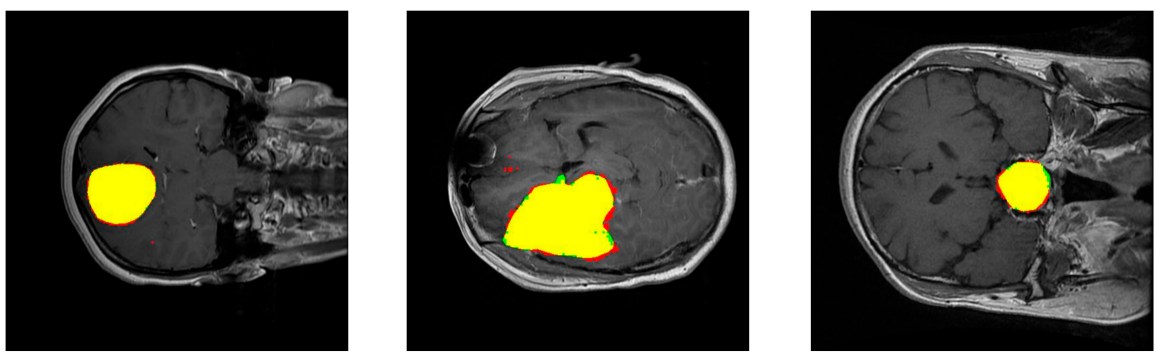

**Figure 4.** Examples of results of the proposed method for three slices corresponding to meningioma, glioma, and pituitary tumors, respectively. The images show the tumor segmentation: The region detected is shown in red while the ground truth region is shown in green. As a result, the intersection region is shown in yellow.

The segmentation metrics obtained are shown in Table 1. The Dice index, which is inversely related to the number of false positives and false negatives extracted (see Equation (3)), is the largest for images with meningioma, then with pituitary tumor, and the lowest with glioma. Therefore, the proportion of false positives and false negatives is the lowest in the prediction of meningiomas and is the largest in the prediction of gliomas. The Sensitivity index, which computes the ratio between true positives and the ground truth (see Equation (4)), led to the same ranking (the largest value for images with meningioma, then with pituitary tumor, and the lowest value with glioma), although the value of this index is closer for images with meningioma and pituitary tumor. The prediction of gliomas had the lowest relation of true positives with respect to the ground truth. On the contrary, the pttas index is the largest for images with glioma, then with pituitary tumor, and the lowest with meningioma. The values of this index, which measures the relation in size between the correctly labeled tumor regions and all the predicted tumor regions in the images, indicates that for gliomas our model misclassified a lower proportion of pixels as belonging to the two other brain tumors than for meningiomas or pituitary tumors. It can be observed that the different features of the three analyzed brain tumors and the terms including in the segmentation metrics led to a different ranking in the segmentation of the three tumors. The values obtained for all the segmentation metrics are remarkable with average values of Dice = 0.828, Sensitivity = 0.940, and pttas = 0.967.

**Table 1.** Segmentation metrics of the 3064 slices processed with 5-fold cross validation.

| Metric\Tumor | Meningioma | Glioma | Pituitary Tumor | Average |
|---|---|---|---|---|
| DICE | 0.894 | 0.779 | 0.813 | 0.828 |
| SENSITIVITY | 0.961 | 0.907 | 0.954 | 0.940 |
| PTTAS | 0.938 | 0.986 | 0.979 | 0.967 |

The histograms for these metrics of the segmentation per processed slice are included in Figure 5. Remarkably, Sensitivity and pttas histograms have their values concentrated on a very narrow interval around the highest zone. This means that almost all the processed images have a high value in those evaluation metrics.

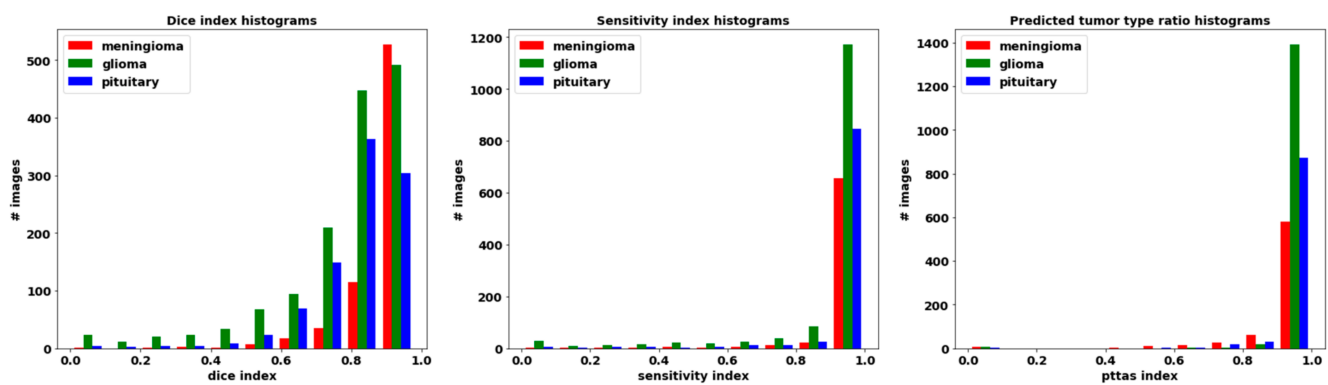

**Figure 5.** Histograms of metrics obtained after processing the dataset. Left: Dice histogram. Center: Sensitivity histogram. Right: Histogram of the predicted tumor type.

Regarding the pttas index histogram (see Figure 5—right), we can see a very outstanding behavior. Around value 0, there are just a few samples. In these samples the predicted tumor is wrong as pttas is lower than 1/3 (pttas measures the ratio to the right label). If pttas is larger than 1/3, it means that the tumor type prediction was correct. The rest of the samples are found in values above 0.4, which means that they have all been correctly predicted. The high values of the Sensitivity index (see Figure 5—center) are due to the fact that almost all the tumor regions have been correctly extracted in the processed images (regions in yellow in Figure 4). A greater amplitude in the Dice histogram (see Figure 5—left) reflects the existence of false positives and false negatives. However, the Dice index results are relatively high, so the number of false positives and false negatives is quite low.

Figure 6 shows some segmentations with misclassified areas; in the example of the upper row, the greater part of the detected tumor is correctly labeled (red area in the right column), resulting in a very correct segmentation (yellow area in the left column). However, there is a region detected in the non-cerebral area wrongly labeled as glioma tumor (green area in the right column). This example shows the added complexity inherent to this dataset due to the fact that it includes non-cerebral areas that can generate false positives. This complexity also manifests itself in the example of the middle row. Similarly, the segmentation is relatively correct (yellow region in the left column), but there is a misclassified area labeled as a pituitary tumor (blue region in the right column) located in the sphenoidal sinuses area, which is where pituitary tumors appear. The physical structure of the sphenoidal sinuses led to confusion to our model. The third example (lower row) shows a confusion between a real glioma (green region in the right column) and a wrongly predicted meningioma region (red area in the left column).

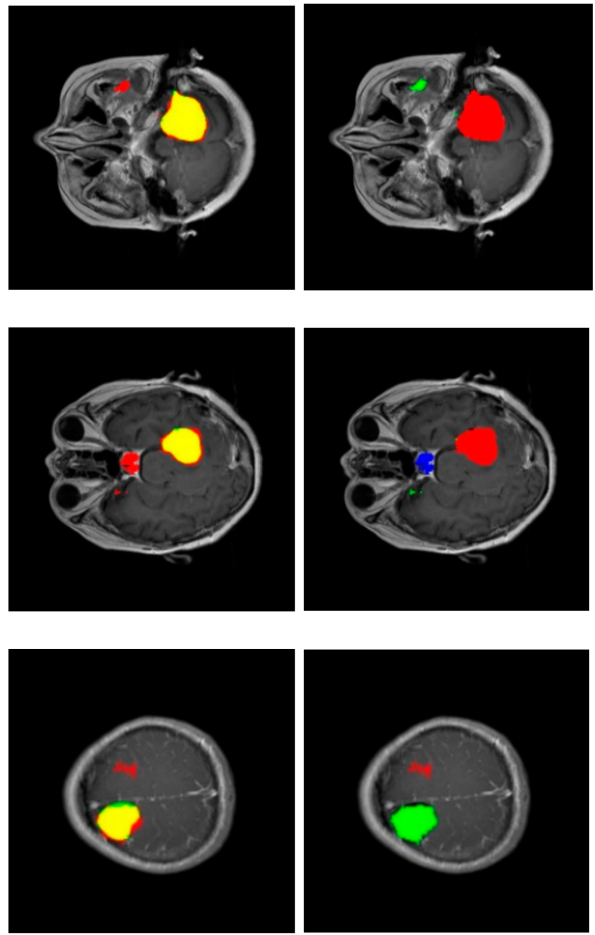

**Figure 6.** Confusion in segmentation. Color-generated images for proper visualization of regions. Upper row corresponds to a meningioma tumor, middle row to a meningioma tumor, and lower row to a glioma tumor. Right column shows the predicted tumor tags (red: meningioma, green: glioma, blue: pituitary tumor). Left column shows segmentations (red: predicted, green: ground truth and, as a result, the intersection is shown in yellow).

In our method, tumor classification is a direct outcome of the resulting segmentation. The classification function for the tumor in the image counts the tumor type prediction for every pixel (pttas, see Equation (5)) and considers the highest value to be the predicted tumor type given that it is greater than the confidence threshold $\tau_c$ (see Equation (2)). Tables 2 and 3 show the values of the confusion matrix and the tumor classification accuracy for Cheng et al. [5] method and our method, respectively. It can be observed that the confusion matrix for our method has very high values in its diagonal, which results in a high tumor classification accuracy of 0.973.

**Table 2.** Confusion matrix of the predicted tumor for Cheng et al. [5] method. The tumor type precision (tumor classification accuracy) achieved ($\sum$ diag/ $\sum$ total) is 0.912.

| True\Predicted | Meningioma | Glioma | Pituitary Tumor | Sensitivity |
|---|---|---|---|---|
| Meningioma | 571 | 57 | 80 | 0.86 |
| Glioma | 44 | 1326 | 56 | 0.96 |
| Pituitary tumor | 75 | 54 | 801 | 0.87 |
| **Tumor classification accuracy** | | | | **0.912** |

**Table 3.** Confusion matrix of the predicted tumor for our method. The number of non-classified images is 61 (42/4/15), taking a confidence threshold $\tau_c = 0.75$. The tumor type precision (tumor classification accuracy) achieved ($\sum$ diag/ $\sum$ total) is 0.973.

| True\Predicted | Meningioma | Glioma | Pituitary Tumor | Sensitivity | Non-Classified |
|---|---|---|---|---|---|
| Meningioma | 659 | 4 | 3 | 0.93 | 42 |
| Glioma | 7 | 1414 | 1 | 0.99 | 4 |
| Pituitary tumor | 1 | 3 | 911 | 0.98 | 15 |
| **Tumor classification accuracy** | | | | | **0.973** |

The classification function, Equation (2), depends on the confidence threshold parameter $\tau_c$. Figure 7 shows the relationship between the confidence threshold and the resulting precision. A slow descent can be observed up to a confidence threshold of 0.88 followed by a sharp decline. There is no strong restriction on the selection of the confidence threshold and, in order to obtain a classification with a precision level higher than 0.9, it could be possible to raise our confidence threshold to a very high value (around 0.92).

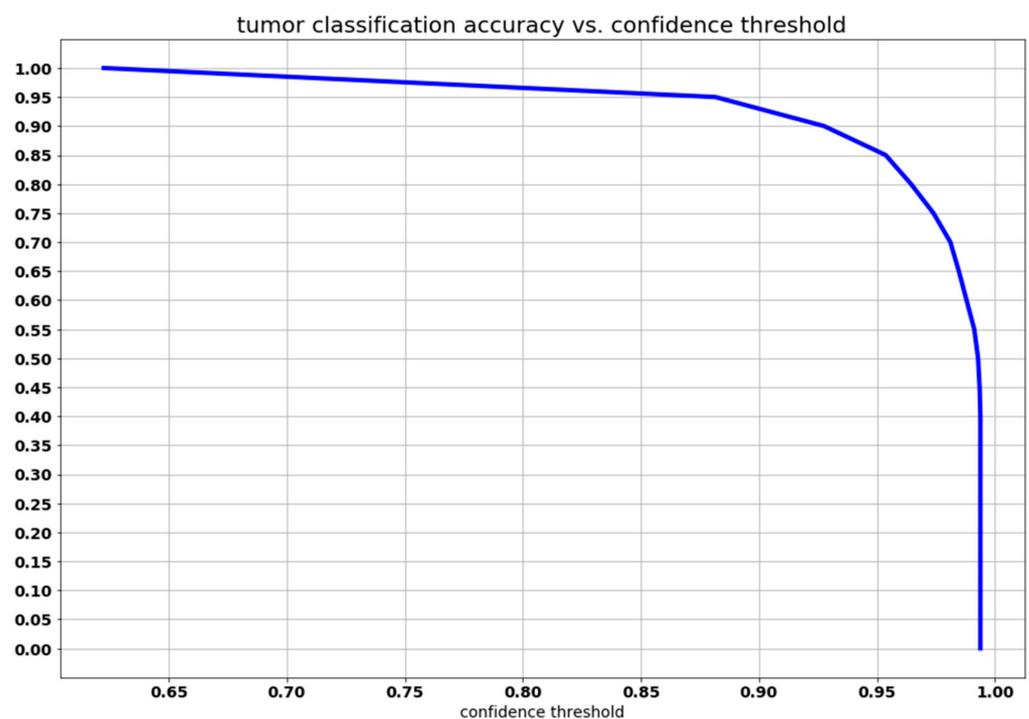

**Figure 7.** Graph of the relation between the confidence threshold and the precision of the predicted tumor type.

This behavior could be predicted observing the pttas index histogram (see Figure 5). Peak values in the histogram can be found once a pttas value of 0.88 is reached. The number of images with a pttas index value between 0.4 and 0.88 is very low, so when the confidence threshold is raised from 0.4 to 0.88, the precision will only be marginally lower as can be observed in the graph in Figure 7.

Due to the fact that all the processed images with a pttas index value greater than 0.4 are correctly labeled, if we were to choose a confidence threshold equal or lower than 0.4, the precision would be close to 100% (all samples would be correctly labeled, except those with a value close to zero). This is why the confidence threshold diagram begins with accuracy ≈1 (= 0.994).

*Comparison with Other Methods*

Segmentation performance metrics of our method (see Table 1) are in the range of the wining proposals in the well-known brain tumor image segmentation challenge BRATS [3]. This benchmark uses glioma tumor images and supplies a dataset with four modalities of MRI (T1, T2, T1-CE, and FLAIR). The 2013 top-ten ranking of the participating methods obtained glioma Dice index values between 0.69 and 0.82 for the segmentation of the complete tumor. The Dice index value obtained by our method for glioma segmentation is 0.779, which is close to the highest achieved value, even with the inconveniences associated with working with a single MRI modality and classifying among three different kinds of tumors.

A more appropriate comparison of our results is with those results obtained by previous works using the same T1-CE MRI image dataset. Table 4 shows the tumor classification accuracy of our approach, together with seven published methods: two feature-driven approaches [5,32] and five deep learning approaches [24–27]. Our method outperforms the other proposals with a tumor classification accuracy of 0.973.

**Table 4.** Comparison of the proposed approach with other approaches over the same T1-CE MRI image dataset.

| Authors | Classification Method | Tumor Classification Accuracy |
|---|---|---|
| Cheng et al. [5] | SVM | 0.912 |
| Cheng et al. [32] | Fisher kernel | 0.947 |
| Abiwinanda et al. [24] | CNN | 0.841 |
| Pashaei et al. [25] | CNN | 0.810 |
| Pashaei et al. [25] | CNN and KELM | 0.936 |
| Sultan et al. [26] | CNN | 0.961 |
| Anaraki et al. [27] | CNN and GA | 0.942 |
| Our approach | Multiscale CNN | 0.973 |

The two studies conducted by Cheng et al. [5,32] are examples of classical machine learning approaches using pre-defined feature extraction. In a first paper, Cheng et al. [5] proposed a classification method based on the technique of splitting the augmented tumor region in ring-shaped sub-regions and extracting local features using three different methods: intensity histogram, gray level co-occurrence matrix (GLCM), and bag-of-words (BoW). Those features were employed to create a dictionary with the K-means technique. Then, a trained classifier was used to classify the ROIs as a tumor type. Three classification methodologies were compared: SVM, sparse representation-based classification (SRC), and k-Nearest Neighbors (kNN). The best results were obtained using BoW and SVM as the classifier. In a second paper, Cheng et al. [32] proposed a content-based image retrieval for MRI images using the Fisher kernel framework. Cheng et al. methodologies [5,32] start with segmentations performed by experienced radiologists, and thus, they are semi-automatic methods. They need an interactive human intervention during classification, whereas our proposed methodology is fully automatic. Nonetheless, using a confidence threshold of 0.75, our method obtains a tumor classification accuracy of 0.973 versus 0.912 in [5] and 0.947 in [32]. Moreover, Figure 7 shows that it is possible to raise the confidence threshold up to 0.9 in order to obtain even a higher classification accuracy. Setting a threshold of 0.75, 61 slices resulted unclassified (2% of the total processed MRI images): 42 meningiomas, 4 gliomas, and 15 pituitary tumors. These figures show to what extent the detection of meningioma tumors is much more demanding. As in Cheng et al. [5] method (see Table 2), in our work the lowest Sensitivity value is for meningioma (see Table 3), due to the lower intensity contrast between tumor areas and healthy areas. Sensitivity for glioma is the highest. Pituitary tumors are somewhere in between meningiomas and gliomas.

The five Deep Learning methods in Table 4 also obtained a lower accuracy than our proposal. Abiwinanda et al. [24] used two convolution layers and 64 fully connected neurons. They did not use the full dataset but selected 700 images from each brain tumor type, with the purpose of balancing the data. A subset of 500 tumor images of each class was used for training and a subset of 200 images for testing. They did not apply data augmentation. Their classification accuracy was 0.841, quite lower than our approach (0.973). Pashaei et al. [25] proposed two Deep Learning methods. In their first method, a CNN model is used for classification with 4 convolution and normalization layers, 3 max-pooling layers, and a last fully connected layer. In their second method, the features extracted by a CNN are the inputs of a KELM classifier. They used 70% of the dataset for training, no data augmentation and 30% for testing using a 10-fold cross validation. The classification accuracy was 0.810 with the first method, and it was significantly improved with the second method (0.936). Sultan et al. [26] used a CNN containing 16 convolution layers, including pooling and normalization, and a dropout layer before the last fully connected layer. They used 68% of the images for training and the remaining 32% for validation and test. Significantly, training data was increased by a factor of 5 with geometric augmentation and grayscale distortion. Their classification accuracy is the second highest in Table 4 (0.961). Finally, Anaraki et al. [27] also proposed a CNN but included a genetic algorithm to select the best configuration of the CNN: number of convolutional, max-pooling and fully connected layers, activation function, dropout parameter, optimization method, and learning rate. Data augmentation was carried out by applying rotation, translation, scaling, and mirroring. They obtained an accuracy of 0.942.

There are some other interesting deep learning approaches that used a different brain tumor MRI dataset in their experiments. Havaei et al. [10] trained several CNN neural networks with the BRATS-2013 glioma tumor dataset. They included a three-step preprocessing: removal of the highest and lowest intensities, intensity inhomogeneity correction, and normalization. The segmentation results were close to those obtained with our method, with a Dice index between 0.77 and 0.88 for the complete tumor and a Sensitivity index between 0.75 and 0.96. Our method obtains a slightly higher Dice index value of 0.779 and a Sensitivity index of 0.907 for glioma segmentation (see Table 1). Mohsen et al. [23] proposed a Deep Learning approach together with DWT and PCA and applied it to classify a dataset of 66 MRIs into 4 classes: normal, glioblastoma, sarcoma, and metastatic bronchogenic carcinoma tumors. They used Fuzzy C-means clustering to separate brain tissues from skull before the deep learning processing. Although the types of brain tumors in the images are different from the ones in the T1-CE MRI image dataset we used, the classifier had to obtain an output label among four different values, similarly to our approach with the adopted T1-CE MRI image dataset. They obtained a classification accuracy of 0.969, lower than our approach.

## 4. Conclusions

In this paper, we present a fully automatic brain tumor segmentation and classification method, based on a CNN (Convolutional Neural Network) architecture designed for multiscale processing. We evaluated its performance using a publicly available T1-weighted contrast-enhanced MRI images dataset. Data augmentation through elastic transformation was adopted to increase the training dataset and prevent overfitting. The measures of performance obtained are in the range of the top ten methods from the BRATS 2013 benchmark. We compared our results with other seven brain tumor classification approaches that used the same dataset. Our method obtained the highest tumor classification accuracy with a value of 0.973. Our multiscale CNN approach, which uses three processing pathways,

 

is able to successfully segment and classify the three kinds of brain tumors in the dataset: meningioma, glioma, and pituitary tumor. In spite of the fact that skull and vertebral column parts are not removed and the variability of the three tumor types, which caused false positives in some images, our approach achieved outstanding segmentation performance metrics, with an average Dice index of 0.828, an average Sensitivity of 0.940, and an average pttas value of 0.967. Our method can be used to assist medical doctors in the diagnostics of brain tumors and the proposed segmentation and classification method can be applied to other medical imaging problems. As future work, we plan to develop an FCN architecture for the classification of the same MRI images dataset and compare its performance with the proposed model. Moreover, we plan to study the applicability of the proposed multiscale convolutional neural network for segmentation in other research fields such as satellite imagery.

**Author Contributions:** Conceptualization, F.J.D.-P.; methodology, F.J.D.-P.; software, M.M.-Z., M.A.-R., and D.G.-O.; experimental results, M.M.-Z., M.A.-R., and D.G.-O.; supervision, F.J.D.-P.; writing—original draft and review, F.J.D.-P., M.M.-Z., M.A.-R., and D.G.-O. All authors have read and agreed to the published version of the manuscript.

**Funding:** This research received no external funding.